\documentclass[aps,prb,amsmath,amssymb,superscriptaddress,notitlepage,twocolumn,reprint]{revtex4-1}
\usepackage{chngcntr}\usepackage{graphicx}
\usepackage{dcolumn}
\usepackage{dirtytalk}
\usepackage{stackengine}
\usepackage[section]{placeins}
\usepackage[dvipsnames*,svgnames]{xcolor}
\usepackage{verbatim}
\usepackage[position=bottom,caption=false,captionskip=3pt,font=normalsize,subrefformat=simple,labelformat=simple]{subfig}

\usepackage{listings}
\usepackage{mathtools}
\usepackage{amsmath}
\usepackage{chngcntr}
\usepackage{bm}
\usepackage{algorithm}
\usepackage{algpseudocode}
\usepackage{url}
\usepackage[super]{nth}
\usepackage{placeins}
\usepackage{nicefrac}    
\usepackage{multirow}
\usepackage[version=3]{mhchem}
\usepackage{siunitx}
\usepackage{microtype}
\usepackage{lipsum}

\usepackage[breaklinks]{hyperref}
\usepackage{tabularx}
\usepackage{soul}
\hypersetup{
colorlinks,
linkcolor={blue!100!black},
citecolor={blue},
urlcolor={blue!80!black}}

\newcommand\varpm{\mathbin{\vcenter{\hbox{%
  \oalign{\hfil$\scriptstyle+$\hfil\cr
          \noalign{\kern-.3ex}
          $\scriptscriptstyle({-})$\cr}%
}}}}
\newcommand\varmp{\mathbin{\vcenter{\hbox{%
  \oalign{$\scriptstyle({+})$\cr
          \noalign{\kern-.3ex}
          \hfil$\scriptscriptstyle-$\hfil\cr}%
}}}}
\usepackage{array}

\usepackage{booktabs}
\def\equationautorefname~#1\null{#1\null}
\usepackage[bottom]{footmisc} \makeatletter \def\p@section{} \def\p@subsubsection{} \makeatother

\begin{document}

\title{Reduced-Order Model to Predict Thermal Conductivity of Dimensionally-Confined Materials}

\author{S. Aria Hosseini}
\affiliation{Department of Chemistry, Massachusetts Institute of Technology, 77 Massachusetts Avenue,
Cambridge, Massachusetts 02139, USA}
\author{Alex Greaney}
\thanks{Correspondence}
\email{greaney@ucr.edu}
\affiliation{Department of Mechanical Engineering, University of California, Riverside, 900 University Avenue, Riverside, CA 92521, USA}

\author{Giuseppe Romano}
\thanks{Correspondence}
\email{romanog@mit.edu}
\affiliation{Institute for Soldier Nanotechnologies, Massachusetts Institute of Technology, 77 Massachusetts Avenue,
Cambridge, Massachusetts 02139, USA}

\begin{abstract}

Predicting nanoscale thermal transport in dielectrics requires models, such as the Boltzmann transport equation (BTE), that account for phonon boundary scattering in structures with complex geometries. Although the BTE has been validated against several key experiments, its computational expense limits its applicability. Here, we demonstrate the use of an analytic reduced-order model for predicting the thermal conductivity in dimensionally confined materials, i.e., monolithic and porous thin films, and rectangular and cylindrical nanowires. The approach uses the recently developed ``Ballistic Correction Model'' (BCM) which accounts for materials' full distribution of phonon mean-free-paths. The model is validated against BTE simulations for a selection of base materials, obtaining excellent agreement. By furnishing a precise yet easy-to-use prediction of thermal transport in nanostructures, our work strives to accelerate the identification of materials for energy-conversion and thermal-management applications.

\end{abstract}

\maketitle

Exploitation of ballistic (non-diffusive) phenomena in thermal transport through dimensionally confined materials has enabled unprecedented control of phonon-mediated heat flow with a wide range of applications, from thermoelectric and photovoltaic generators,~\cite{qiao2019tailoring, reddy2019review, PhysRevB.102.205405, de2016cadmium, jackson2016effects, zhang2015flexible}  through electronic and optoelectronic devices,~\cite{scott2018thermal, heiderhoff2017thermal, kim2014wearable} to thermal management and thermal cloaking~\cite{xiao2021inverse, burton2018thin, chowdhury2009chip}. It has also been taken advantage of in ferroelectric memories and electrostatic super-capacitors~\cite{park2018review, park2015toward}.  While reliable models exist to compute the thermal conductivity of bulk materials, predicting the effective thermal conductivity, $\kappa_{\mathrm{eff}}$, of nano-/micro- scales films and wires is more challenging.~\cite{kerdsongpanya2017phonon, cahill2014nanoscale} As a result, most computational studies of these structures use semi-analytical models,~\cite{minnich2015thermal, vermeersch2016cross}  or numerically solve the spatially resolved Boltzmann transport equation (BTE) either stochastically,~\cite{HOSSEINI2022100719} or deterministically~\cite{harter2019prediction, romano2021openbte}.

While significant progress has been made, these approaches' complexity and computational cost highlight the need for robust reduced order models that can reliably predict the effective thermal conductivity, $\kappa_{\mathrm{eff}}$, of dimensionally confined materials. To meet this need we recently introduced the ``Ballistic Correction Model'' (BCM) ~\cite{hosseini2022universal} which provides a simple closed form analytic prediction of  $\kappa_{\mathrm{eff}}$ that can be used to quickly screen material/geometry permutations for a desired application. The BCM is built on two key approximations: (1) It accounts for the contributions to transport across the full phonon spectrum by approximating the cumulative conductivity as a logistic function. (2) It uses Matthiessen's rule with an appropriately chosen length-scale to approximate the suppression of conductivity across the phonon spectrum due to the system geometry. In a prior work, we established the validity of the first of these approximations across a range of Group-IV, and III-V dielectrics, along with their binary and ternary alloys.~\cite{HOSSEINI2022123191} Here we present a deeper examination of the second approximation.

In the remainder of this letter, we first concisely reintroduce the BCM as developed to model materials with nanoscale porosity, and demonstrate that with the appropriate choice of characteristic length-scale it can also predict $\kappa_{\mathrm{eff}}$ along and across thin films.  We then apply the BCM to nanoporous films and rectangular and cylindrical nanowires which have two different and orthogonal length-scales of confinement. We finish this letter by briefly comparing alternative strategies for choosing the appropriate geometric length-scale, and identify whether length-scales can be reliably combined using Matthiessen's rule or via a multiplicative rule that was suggested in Ref~\cite{hao2020two}. Our results highlight some inherent limitations of Matthiessen's rule which are also briefly discussed.

\begin{figure*}[t]
\centering
\includegraphics[width=1\textwidth]{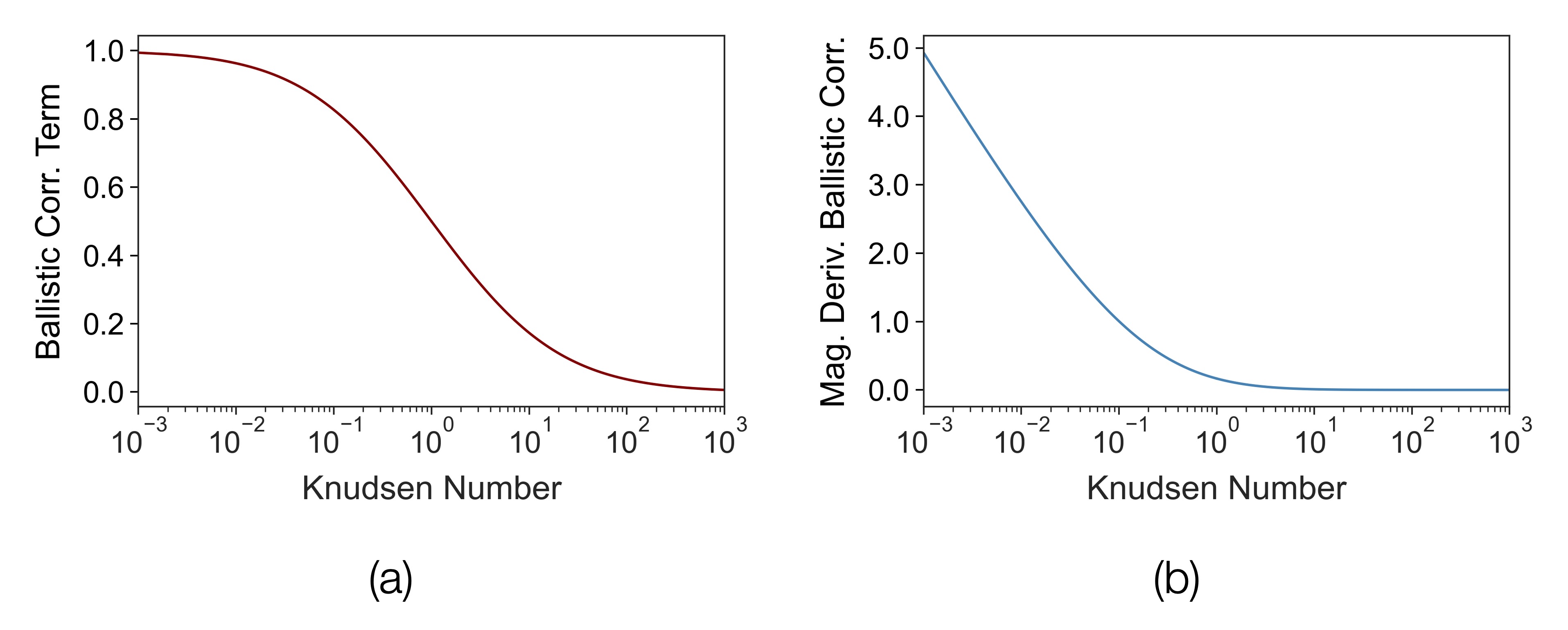}
\caption{(a) The ballistic correction term $\Xi$ and (b) its first-order derivative vs. the Knudsen number. The scale-dependent correction term is considerably small in the ballistic regime where $\mathrm{Kn} \gg 1$. Small changes in $\mathrm{Kn}$ lead to significant reduction in $\Xi$ in the near-diffusive regime, however $\Xi$ is quite insensitive to further engineering materials to the tail of the ballistic regime.}
\label{fig:Kn}
\end{figure*}

The Ballistic Correction Model is based on the recognition that the reduced $\kappa_{\mathrm{eff}}$ of a nanostructured material arises from suppression of phonons across the full spectrum of phonon mean free paths (MFPs). Mathematically, this can be expressed with the integral $\kappa_{\mathrm{eff}} = \int_0^{\infty} K(\Lambda) S(\Lambda)  d\Lambda$, where $K(\Lambda)$ is the contribution to the thermal conductivity of the bulk materials from phonons with intrinsic MFP $\Lambda$, and $S(\Lambda)$ is the suppression function which describes how the contribution from these phonons is suppressed in the nanostructured geometry.~\cite{yang2013mean} The crux of the BCM's approach is to replace $K(\Lambda)$ and $S(\Lambda)$ with reasonable algebraic approximations that permit this integral to be performed analytically---and thus the validity of the BCM hinges on accuracy of the algebraic approximations in representing these terms.

The first BCM approximation is to represent the normalized cumulative distribution of thermal conductivity over the MFP in the bulk material, $\alpha(\Lambda)$, with a logistic function on a log scale abscissa so that $\alpha(\Lambda) = \frac{1}{\kappa_{\mathrm{bulk}}}\int_0^\Lambda K(\Lambda') d\Lambda'\approx \left[1+\frac{\Lambda_o}{\Lambda}\right]^{-1}$, where the characteristic MFP, $\Lambda_o$, is a fitting parameter that describes the median MFP of the conductivity distribution. It results in the approximating function $K(\Lambda)=\frac{\Lambda_o}{\left(\Lambda+\Lambda_o\right)^2}$. This logistic function approximation has been shown to provide a very good practical approximation of the heat transport spectrum in many bulk dielectric materials,~\cite{hosseini2022universal} but its suitability for representing 2D materials has not yet be tested. To facilitate use of the BCM, the BCM's GitHub repository~\cite{Hosseini2021} includes a public database of the cumulative thermal conductivity for a wide range of  materials computed from first principles, along with their logistic fits.

The suppression function due to phonon-boundary scattering is approximated with $S(\Lambda)\approx S(0)\left[1+\Lambda/L_w \right]^{-1}$. This has its basis in Matthiessen's rule, with $L_w$ being the weighted average line-of-sight distance to a boundary or other geometric feature. The term $S(0)$ is the  macroscopic suppression function, which is equal to $\nicefrac{\kappa_{\mathrm{f}}}{\kappa_{\mathrm{bulk}}}$, where $\mathrm{\kappa_{f}}$ is the Fourier's law prediction of the effective thermal conductivity and $\kappa_{\mathrm{bulk}}$ is the bulk thermal conductivity.~\cite{hosseini2022universal} The two parameters $L_w$ and $S(0)$ depend on the system geometry but are material independent.

With these two approximations for $K(\Lambda)$ and $S(\Lambda)$, the BCM yields the effective thermal conductivity
\begin{equation}\label{eq:keff}
\kappa_{\mathrm{eff}} = \kappa_{\mathrm{bulk}}S(0)\Xi(\mathrm{Kn}),
\end{equation}
where $\Xi(\mathrm{Kn})$ is the ballistic correction term which accounts for truncation of long MFP phonons and is given by
\begin{equation}\label{eq:Xi}
    \Xi (\mathrm{Kn}) = \left[\frac{1 + \mathrm{Kn} \left( \ln(\mathrm{Kn}) - 1  \right)}{\left ( \mathrm{Kn} - 1 \right)^2} \right].
\end{equation}
with the Knudsen number $\mathrm{Kn}=\nicefrac{\Lambda_o}{L_w}$.

For $\mathrm{Kn} = 1$, we have $\kappa_{\mathrm{eff}} = \nicefrac{1}{2}\kappa_{\mathrm{f}}$. For $\mathrm{Kn} \rightarrow 0$, $\Xi \rightarrow 1$, recovering the diffusive regime. For large $\mathrm{Kn}$, i.e., in the ballistic regime, $\Xi(\mathrm{Kn}) \approx \ln{(\mathrm{Kn})} \mathrm{Kn}^{-1}$. The first-order derivative of Eq.~\eqref{eq:Xi} is 
\begin{equation}\label{eq:dXi}
    \Xi' \left ( \mathrm{Kn} \right ) = -\left[\frac{\left ( \mathrm{Kn}+1 \right )\ln\left ( \mathrm{Kn} \right )-2\left (\mathrm{Kn}-1\right)}{\left ( \mathrm{Kn}-1 \right )^3}\right].
\end{equation}
Figures~\ref{fig:Kn}(a\&b) show the ballistic correction term and the magnitude of its first-order derivative versus the Knudsen number. While there is a large scope to reduce thermal conductivity in the tail of the ballistic regime ($\mathrm{Kn} >> 1$), the derivative of the correction term, $\Xi'$, is extremely small in this regime, implying that any effort to further push the transport to the Knudsen regime has a negligible effect on reducing thermal conductivity. 

\begin{figure*}[t]
\centering
\includegraphics[width=1\textwidth]{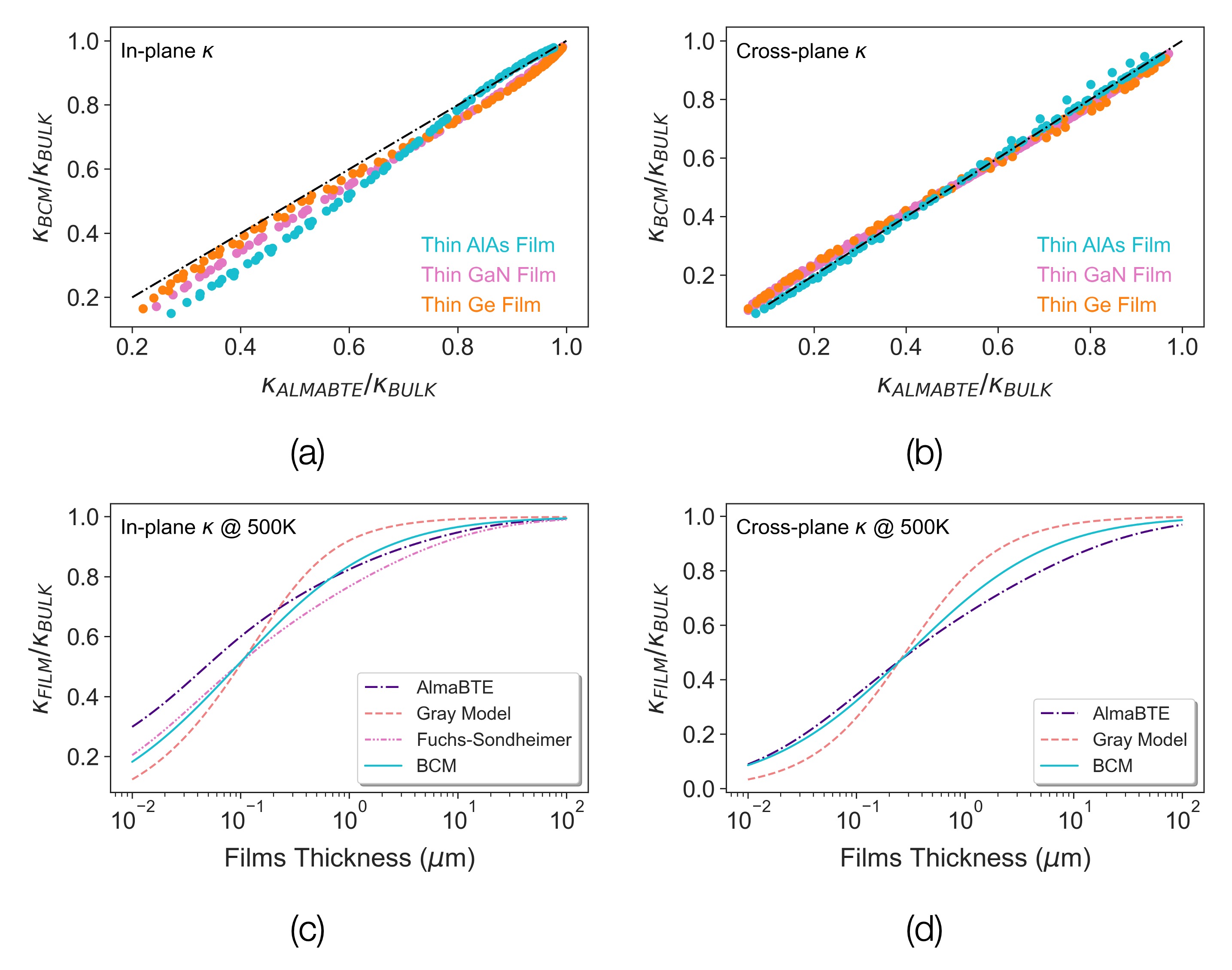}
\caption{Thermal conductivity reduction predicted by the reduced-order model vs. semi-analytical simulations performed in AlmaBTE for a selection of three base materials (Ge, GaN, and AlAs) at a variety of different temperatures and thicknesses for (a) in-plane (b) cross-plane thermal conductivities. The thickness varies from 10 nm up to 10 micron and the temperature varies from 300--800 K. Plots (c) and (d) show thermal conductivity reduction in Si at 500 K for the in-plane and cross-plane phonon transports, respectively. BCM is closed-form, easy to use, and outperforms gray models via counting for the phonons MFP distributions.}
\label{fig:2}
\end{figure*}

In our prior works we showed that the BCM as presented above could accurately predict thermal conductivity reduction in materials with nanoscale pores.~\cite{hosseini2022universal, HOSSEINI2022123191} In the following few paragraphs we apply BCM to predict in-plane and cross-plane thermal conductivities of thin films.

The directionally averaged suppression functions for phonons transporting heat along or across a thin slab can be derived starting from the mode-resolved suppression functions
\begin{equation}\label{eq:moderesolved}
\begin{matrix}
 S_\lambda^{\parallel} = 1-\frac{\Lambda_\lambda^\zeta}{L}\left[1-\exp\left({-\frac{L}{\Lambda^\zeta_\lambda}}\right)\right], \\
S_\lambda^{\perp} =  \left[1+2\frac{\Lambda^\zeta_\lambda}{L} \right]^{-1},
 \end{matrix}
\end{equation}
for the in-plane~\cite{minnich2015thermal} and cross-plane~\cite{vermeersch2016cross} transport, respectively. The term $\Lambda_\lambda^\zeta$ is the mode-resolved vectorial MFP projected along the film surface normal $\zeta$, and $L$ is the thickness. The vectorial MFPs can be highly anisotropic with long MFPs along high symmetry crystal axes while still satisfying von Neumann's principle that the net tensor of the bulk crystal be isotropic.~\cite{romano2021,hurley1985phonon} Within the relaxation time approximation, the effective thermal conductivity is $\kappa_{\mathrm{eff}} = \sum_\lambda \kappa_{\zeta \zeta}^\lambda S_\lambda$, where $\kappa^\lambda_{\zeta\zeta}$ is the mode-resolved bulk thermal conductivity projected along $\zeta$. If, for simplicity, one neglects the anisotropy in the MFP distribution and assumes it to be isotropic, Eqs.~\eqref{eq:moderesolved} reduce to~\cite{vermeersch2016cross}
\begin{equation}\label{eq:moderesolved2}
\begin{matrix}
S_{\parallel}(\Lambda) = 1- \frac{3}{8}\mathbb{K}\left[ 1-4\left(\mathbb{E}_3(\mathbb{K}^{-1})-\mathbb{E}_5(\mathbb{K}^{-1})\right)\right], \\ 
S_{\perp}(\Lambda) =  \frac{3}{8\mathbb{K}^3}\left[ 2\mathbb{K} (\mathbb{K}-1) + \ln(1+2\mathbb{K})\right],
\end{matrix}
\end{equation}
where $\mathbb{K} = \nicefrac{\Lambda}{L}$, and $\mathbb{E}_n(\eta) = \int_{0}^{1}\mu^{n-2}\exp(\nicefrac{-\eta}{\mu})d\mu$.~\cite{chen2005nanoscale}

The suppression term $S_{\parallel}$ in Eqs.~\eqref{eq:moderesolved2} is the exact solution of BTE for isotropic gray population of phonons and is known as Fuchs-Sondheimer model.~\cite{minnich2015thermal, chen2005nanoscale} Following the BCM, we approximate the suppression functions in Eqs.~\eqref{eq:moderesolved2} with a logistic curve $S(\Lambda) = \nicefrac{1}{1+\frac{\Lambda}{L_w}}$, obtaining $L_w=2.33L$ and $L_w=0.70L$ for the in-plane and cross-plane transports, respectively. These values are slightly lower than those found by Majumdar for gray phonons,~\cite{10.1115/1.2910673} i.e., effective lengths of $L_w=\nicefrac{8}{3}L$ for the in-plane transport and $L_w=\nicefrac{3}{4}L$ for the cross-plane transport. To be precise, the exact characteristic length-scale is not scale-independent but varies non-monotonically with the Knudsen number. The solution by Majumdar is reliable in the diffusive regime, but in the diffusive to ballistic crossover regime a better prediction is obtained using the slightly shorter characteristic lengths suggested in this study. It should be noted that although approximating Eqs.~\eqref{eq:moderesolved2} with logistic forms may introduce modest erroneous in the exact solution, it enables the derivation of the simple analytical solution of Eq.~\eqref{eq:keff}. This is particularly useful in cases such as high-throughput screening, where computational efficiency or simplicity is a priority.

BCM predictions for in-plane and cross-plane transports are validated against Eqs.~\eqref{eq:moderesolved} solved using AlmaBTE~\cite{carrete2017almabte}. The phonon dispersions and the scattering rates were computed on at least a $30\times30\times30$ points Brillouin zone mesh; the second and third-order interatomic force constants for bulk materials, computed with density functional theory and using the virtual crystal approximation, were obtained from the AlmaBTE materials database.~\cite{carrete2017almabte}

Figures~\ref{fig:2}(a\&b) show the in-plane and cross-plane conductivities of AlAs, GaN, and Ge films predicted by the BCM against the BTE prediction. The comparison is performed for films with thicknesses from 10 nm to 10 microns and temperatures from 300 up to 800 K. These plots corroborate the accuracy of the BCM in both diffusive and ballistic regimes. To further examine the validity of BCM, we plot the thermal conductivity reduction of thin Si films at 500 K ($\mathrm{\Lambda_o = 212\ nm}$)~\cite{Hosseini2021} versus thickness in Figs.~\ref{fig:2}(c\&d). The BTE solutions computed using AlmaBTE are represented by the purple curves in these figures. The orange curves are gray models, that is, using median MFP of Si at 500 K ($\Lambda_o$ in BCM) in Fuchs-Sondheimer and Majumdar equations to compute the reduction in thermal conductivity for a gray population of phonons. Note that gray models are often used for their simplicity, but, the BCM is able to capture the full spectral range of mean free paths while still maintaining its simplicity. The pink curve in Fig.~\ref{fig:2}(c) is the isotropic Fuchs-Sondheimer model. We remark that unlike BCM, Fuchs-Sondheimer model is not an stand-alone formula and the bulk mean free path distribution is needed for computing thermal conductivity in nanostructures. In Fig.~\ref{fig:2}(c), the MFP spectral computed using AlmaBTE has been used. The blue curves are BCM solutions. This model gives a good prediction in the regime where Knudsen number is larger or close to unity (relatively large thicknesses). In the Knudsen regime (i.e., for small thicknesses), however, the error becomes appreciable for in-plane transport. In this case the BCM converges to the isotropic Fuchs-Sondheimer model. This effect is due to there being less internal phonon isotropization in this regime, and thus anisotropy in the MFP distribution makes the orientation of the slab relative to the crystal axes matter. The assumption in the BCM that the MFP distribution is isotropic is currently the largest limitation of the model and future work is planned to incorporate orientation dependence and anisotropy.

Next, we use Eq.~\eqref{eq:keff} for the in-plane transport in nanoporous thin film as an example of structures with two degrees of confinement. To identify the two dominant length-scales in nanoporous thin films, we first examine nanoporous materials and thin films, independently. We begin with bulk materials with nanoscale porosity (in the absence of boundary scattering with the film's boundary). The characteristic MFP and suppression function for pores with different shapes and volume fractions are tabulated in Ref~\cite{hosseini2022universal}. The second feature size is due to scattering at the boundary of the thin films. We refer to the first half of this paper, for BCM in thin films.

\begin{figure*}[t]
\centering
\includegraphics[width=1\textwidth]{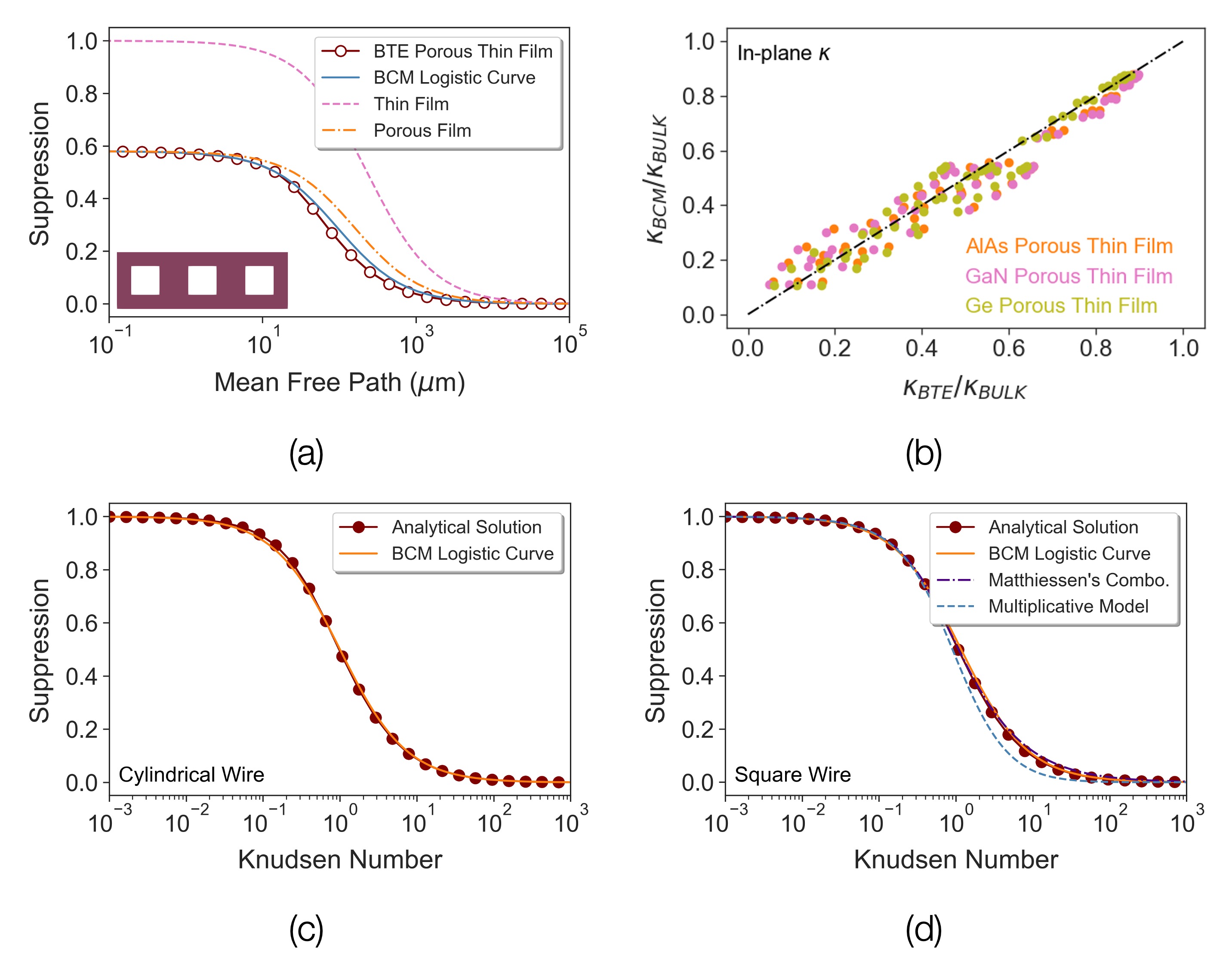}
\caption{(a) The Suppression function of in-plane phonon transport in 100 $\mu$m thin film with 0.25 volume fraction porosity (shown in inset). The maroon curve with open circles is the BTE prediction and the blue solid line shows the logistic curve. The MFP suppression by porosity is shown in the dot-dash line and the suppression by boundary scattering at the surface of the thin film is shown with a dashed line. (b) The in-plane thermal conductivity reduction predicted by the reduced-order BCM vs. full BTE simulations for a selection of three base materials (AlAs, GaN and Ge) with nanoscale pores with two different porosities of 0.05 and 0.25 and a variety of spacings. The periodicity varies from 10 nm up to 10 micron and the temperature varies from 300--800 K. (c\&d) Suppression functions predicted by the reduced-order BCM (solid orange lines) vs. BTE analytical solutions (red curves with circle markers) for nanowires with circular and square cross-sections. The indigo dot-dash and blue dash curves in the latest pane are Matthiessen’s and multiplicative combination of gray-BTE solution for slabs assuming independent orthogonal axes of confinements, respectively.}
\label{fig:3}
\end{figure*}

Fig.~\ref{fig:3}(a) shows the suppression function for in-plane transport in nanoporous thin film. In this figure the BTE prediction of a nanoporous thin film in the material diffusive limit~\cite{hosseini2022universal} is plotted in maroon with open circle marker. The film thickness, $L$, is set to 100 microns. In this case, the suppression functions are calculated by first interpolating the mode-resolved suppression function in angular and MFP space, followed by an angular average.~\cite{romano2021,romano2017directional}  We also plotted the suppression functions induced by the boundary scattering of the thin film and scattering at the surface of the pores in dash-line and dot-dash line, respectively. The blue solid curve is the estimation by the logistic curve and effective length from the Matthiessen's rule for the same material, i.e., for structures with feature sizes $L_i$ ($i=1,2$), the effective length is $L_w^{-1} = \sum_iL_i^{-1}$, and thus $\mathrm{Kn} = \sum_i\mathrm{Kn}_i$. The term $L_1$ is the effective length of the film in the absence of porosity and is equal to $2.33L$ for the in-plane transport (233 micron in this case). The term $L_2$ is the characteristic length due to the porosity. As mentioned above, the magnitude of the $L_2$ depends on the shape and volume fraction of the pores and can be found in Ref~\cite{hosseini2022universal}. For the prismatic pores with square cross-section and $\phi = 0.25$, $L_2$ is equal to $1.58L$ (158 micron in this case). Therefore, the predicted effective length for the porous thin film is 94 micron. We remark that effective length computed by fitting a logistic curve to suppression function from BTE is about 84 micron which is slightly less that the value predicted by BCM. This figure demonstrates that the logistic curve in BCM is robust in providing a good approximation of the suppression function in nanostructures with two degrees of confinement.

We employ Eq.~\eqref{eq:Xi} to compute the reduction in thermal conductivity of porous thin films. We have compared BCM prediction of the in-plane thermal conductivities in AlAs, GaN and Ge porous thin films against a set of BTE simulations in OpenBTE~\cite{romano2021openbte} in Fig.~\ref{fig:3}(b). In these calculations the pores were prismatic with square cross-section, and the spacing between pores varied from 10 nm up to 10 micron and the temperature varied between 300 K and 800 K. Two different porosities of $\phi = 0.05$ and $\phi = 0.25$ were considered. The macroscopic suppression $S(0)$ for these two porosities are 0.90 and 0.56, respectively.~\cite{hosseini2022universal} The BCM is in a good agreement with BTE predictions and thus can be used to compute thermal conductivity in thin films with and without porosity.

To further advance our discussion, we proceed to employ the BCM to model nanowires. The analytical solution to the Boltzmann transport equation for the suppression function in wires with circular cross-section is ~\cite{Sondheimer_1952, doi:10.1098/rspa.1950.0077, 10.1115/NANO2005-87063}
\begin{equation}\label{eq:c_wires}
S(\Lambda)  = 1 - \frac{12}{\pi} \int_{0}^{1}\left ( 1-\mu^2 \right )^{\frac{1}{2}} \Im (\nicefrac{\mu}{\mathbb{K}})d\mu,
\end{equation}
where $\Im (\eta ) = \int_{1}^{\infty } \exp\left (-\eta u\right ) \left ( u^2 - 1  \right )^{\frac{1}{2}} u^{-4} du$. In Eq.~\eqref{eq:c_wires}, the term $\mathbb{K}$ is the phonon MFP normalized by the wire's cross-section diameter $D$. A logistic curve is fitted to Eq.~\eqref{eq:c_wires}, resulting in an effective length of $L_w = D$. Yang and Dames suggested similar logistic form for the suppression function in cylindrical wires.~\cite{yang2013mean} The errors in this approximation never exceed 6\% compared to the exact solution given by Dingle in Ref~\cite{doi:10.1098/rspa.1950.0077}. Yang obtained $\nicefrac{\kappa_{\mathrm{wire}}}{\kappa_{\mathrm{bulk}}}$ values of 0.27, 0.56, and 0.88 for wire diameters of 0.1$\mu$m, 1$\mu$m, and 10$\mu$m from first-principles calculations, respectively. BCM readily returns 0.26, 0.62, 0.89 for similar calculations. The variation of the suppression function with Knudsen number in circular nanowires and the logistic curve are shown in Fig.~\ref{fig:3}(c). 

The analytical solution for nanowires with square cross-section has been derived by MacDonald and Sarginson.~\cite{doi:10.1098/rspa.1950.0136} The final form of the solution is extensive so rather than reproduce it here we refer the reader to the original work. Remarkably, using a logistic form of suppression function with the effective length of $L_w = \sum_iL_i^{-1}$ (and thus $\mathrm{Kn} = \sum_i\mathrm{Kn}_i$) is in very good agreement with the analytical solution. We note that here $L_1 = L_2 = 2.33 L$, where $L$ is the side length of the square (see derivation for in-plane transport in thin films). We conclude that $L_w \approx 1.15L$ for square wires. It is also worth mentioning that deep in the Knudsen regime the analytical solution leads to $S = 1.12\mathbb{K}$, where $\mathbb{K} = \nicefrac{\Lambda}{L}$. This agrees well with the value logistic form yields. The analytical solution given by MacDonald and the logistic form are compared in Fig.~\ref{fig:3}(d). Overall, we observe an excellent agreement between logistic forms (orange curves) and analytical solutions (red curves with solid markers) for both circular and square wires. Finally, it should be noted that in the case of nanowires with rectangular cross-section, the simple form of $\mathrm{Kn} = \sum_i\mathrm{Kn}_i$ is generally applicable in most instances. Nevertheless, in cases where both $\mathrm{Kn}_i$ are within the Knudsen regime, this model may deviate slightly from the analytical solution. This deviation from Matthiessen's combination of slabs in the ballistic regime is understandable given that this is the regime where nanowires, with two degrees of confinement, are furthest from slabs with only one degree of confinement. Further discussion on nanowires with rectangular cross-section will be presented in future works.

We conclude our study by a brief discussion on the choices of the suppression function for structures with multiple degrees of confinement other than the Matthiessen's combination. Let us assume we have a structure with orthogonal feature sizes $L_1$ and $L_2$. An alternative form of the suppression function may have the form of $S(\Lambda) \approx S(0)\left[1+\frac{\Lambda}{L_1}\right]^{-1}\left[1+\frac{\Lambda}{L_2}\right]^{-1}$. This multiplicative form of suppression was first proposed in Ref~\cite{hao2020two}. Following the same procedure as for the BCM with single length-scale, the thermal conductivity reduction will be $\Xi =
\frac{\mathrm{Kn}_{1}^2\ln \mathrm{Kn}_{1}}{\left ( \mathrm{Kn}_{1}-1 \right )^2\left ( \mathrm{Kn}_{1}-\mathrm{Kn}_{2} \right )}
+\frac{\mathrm{Kn}_{2}^2\ln \mathrm{Kn}_{2}}{\left ( \mathrm{Kn}_{2}-1 \right )^2\left ( \mathrm{Kn}_{2}-\mathrm{Kn}_{1} \right )} 
+\frac{1}{\left ( \mathrm{Kn}_{1}-1 \right )\left ( \mathrm{Kn}_{2}-1 \right )}$,
with $\mathrm{Kn}_{1} = \nicefrac{\Lambda_o}{L_1}$ and $\mathrm{Kn}_{2} = \nicefrac{\Lambda_o}{L_2}$. In the limit that either $\mathrm{Kn}_{1}$ or $\mathrm{Kn}_{2}$ goes to 1, this becomes $\Xi=\frac{2 \mathrm{Kn}^2 \ln\left[\mathrm{Kn}\right]+\mathrm{Kn}(4-3\mathrm{Kn})-1}{2 (\mathrm{Kn}-1)^3}$, where $\mathrm{Kn}$ is the non-unitary Knudsen number and in case $\mathrm{Kn}_1 = \mathrm{Kn}_2 = 1$, we will have $\kappa_\mathrm{eff} = \nicefrac{1}{3}\kappa_\mathrm{f}$. In the limit that $\mathrm{Kn}_1\rightarrow \mathrm{Kn}_{2} = \mathrm{Kn}$, we have $\Xi=\frac{\mathrm{Kn}^2-2 \mathrm{Kn} \ln\left[\mathrm{Kn}\right]-1}{(\mathrm{Kn}-1)^3}$. Finally, in the limit that $\mathrm{Kn}_1\rightarrow 0$ or $\mathrm{Kn}_{2}\rightarrow 0$, this equation reduces to Eq.~\eqref{eq:Xi}. The blue dash curve and the indigo dot-dash curve in Fig.~\ref{fig:3}(d) represent the suppression functions for the square nanowires, obtained using the multiplicative and Matthiessen's combinations of Fuchs-Sondheimer model, respectively. It should be noted that, unlike the logistic form where the characteristic length was computed using Matthiessen's rule, in these curves the exact solutions of BTE for the suppression function were combined. This is $S(\Lambda) = S_{\mathrm{FS}}^2$ and $S(\Lambda) = \frac{S_{\mathrm{FS}}}{2-S_{\mathrm{FS}}}$ for the multiplicative and Matthiessen's combinations, respectively. The logistic curve and Matthiessen's combination of the suppression function exhibit a high degree of similarity. However, the multiplicative approach is only accurate in the diffusive regime, as it fails to capture the suppression of thermal conductivity in the diffusive-to-ballistic crossover regime. This indicates that although Matthiessen’s rule is not perfect, it is more reliable than the alternative method examined here and can serve as a useful rule of thumb for selecting the optimal length-scale in similar confined materials.

To summarize, we have assessed the validity of our reduced-order “Ballistic Correction Model”~\cite{hosseini2022universal} for predicting the thermal conductivity of dimensionally confined materials. We have further evaluated the model's accuracy in capturing the reduction in thermal conductivity in materials with multiple degrees of confinement. The model provides accurate estimation of in-plane and cross-plane transport for a wide range of nanomaterials including thin films, films with nanoscale porosity, and nanowires that span both the ballistic and diffusive regimes. We also investigated the inherent limitations of using Matthiessen’s rule for modeling the phonon suppression in confined geometries. It is well understood from exact solutions to the grey-BTE that the characteristic length-scale needed to represent the suppression using Matthiessen’s rule is not scale independent but varies non-monotonically with Knudsen number. Typically the leading scale-independent term from the exact solution is used, but we showed that a more accurate prediction can be achieved in the diffusive to ballistic crossover regime by using a slightly shorter characteristic length. By analyzing rectangular and cylindrical wires for which there are exact but complex solutions to the BTE, we explore different strategies for selecting the appropriate geometric length-scale and investigate whether length-scale can be combined using either Matthiessen’s or multiplicative rules. Our findings suggest that while neither approach is perfect, Matthiessen’s combination is more reliable than the alternative examined. Finally, we proposed a useful rule of thumb for selecting the optimal length-scale in a variety of confined materials. In conclusion, with an appropriate choice of characteristic length-scale, BCM provides a simple yet accurate prediction for the effective thermal conductivity of nanostructured materials that captures all of the ballistic and non-grey spectral behavior of the phonon population and can facilitate the design and discovery of materials for thermal-related and energy-conversion applications without the need to perform complex Boltzmann transport simulations.

The data that support the findings of this study are available from the corresponding author upon reasonable request.

The authors declare no conflict of interest.

This article may be downloaded for personal use only. Any other use requires prior permission of the author and AIP Publishing. This article appeared in S. A. Hosseini et al., Appl. Phys. Lett. 122 (26): 262202 (2023) and may be found at \url{https://doi.org/10.1063/5.0149792}.

\FloatBarrier
\bibliography{references} 

\end{document}